\documentclass[10pt]{article}
\usepackage{conf_iap,psfig}
\begin{document}

\heading{High-velocity clouds: a diverse phenomenon}
\par\medskip\noindent
\author{B.P. Wakker $^{1}$}
\address{University of Wisconsin, 475 N Charter St, Madison, WI53706, USA}

\def\HI{H\,I}
\def\CIV{C\,IV}
\def\NI{N\,I}
\def\NII{N\,II}
\def\NV{N\,V}
\def\OI{O\,I}
\def\OVI{O\,VI}
\def\AlII{Al\,II}
\def\SiII{Si\,II}
\def\SII{S\,II}
\def\FeII{Fe\,II}
\def\dex#1{10$^{#1}$}
\def\tdex#1{$\times$10$^{#1}$}
\def\cmm#1{cm$^{-#1}$}
\def\kms{km\,s$^{-1}$}
\def\deg{$^\circ$}
\def\fdeg{$^\circ$}
\def\Msun{M$_\odot$}
\def\l{$\lambda$}
\def\ll{$\lambda\lambda$}

\begin{abstract}
In this contribution the current state of knowledge about the high-velocity
clouds (HVCs) is summarized. Recent progress has shown that the HVCs are a
diverse phenomenon. The intermediate-velocity clouds (IVCs) are likely to be
part of a Galactic Fountain. The Magellanic Stream is a tidal remnant. HVC
complex C (possibly complexes A and GCN) are low-metallicity clouds near the
Galaxy; they could be remnants of the formation of the Galaxy or old tidal
streams extracted from nearby dwarf galaxies. Having a substantial number of
\HI\ HVCs dispersed throughout the Local Group seems incompatible with the
observed \HI\ mass function of galaxies. Finally, FUSE finds high-velocity \OVI,
some of which is clearly associated with \HI\ HVCs, but some which is not.
\end{abstract}
\section{Introduction}
The HVCs and IVCs consist of gas moving at velocities incompatible with a simple
model of differential galactic rotation. Wakker\cite{W91} defines the
``deviation velocity'' (v$_{\rm dev}$) as the difference between the observed
LSR velocity and the maximum velocity that can be understood in terms of
galactic rotation. With this definition HVCs have $\vert$v$_{\rm
dev}$$\vert$$>$$\sim$90\,\kms, while IVCs have $\vert$v$_{\rm
dev}$$\vert$=30--90\,\kms.
\par Maps of the large-scale structure of HVCs\cite{WABS} in 21-cm \HI\ emission
are based on four surveys. Two of these\cite{Baj,HWS}, cover the sky on a
1\deg$\times$1\deg\ grid (2\deg$\times$2\deg\ for declinations $<$$-$18\deg),
down to a column density of $\sim$2\tdex{18}\,\cmm2, though only at 16\,\kms\
velocity resolution. The Leiden-Dwingeloo Survey (LDS\cite{LDS}) and its
southern equivalent\cite{Arnal,Morras} cover the sky on a 0\fdeg5$\times$0\fdeg5
grid, at 1\,\kms\ velocity resolution, but only down to about 8\tdex{18}\,\cmm2.
\par Most of the recent progress in understanding the origin of the HVCs and
IVCs has been made through measurements of distances and metallicities.
Wakker\cite{WABS} summarizes the available literature up to mid 2001. Below, a
short summary is given of each of the origins for which observational evidence
now exists.


\section{Intermediate-Velocity Clouds}
Distance measurements exist for several IVCs: the IV-Arch ($z$=0.7--1.7 kpc),
the LLIV-Arch (z=0.9 kpc), complex~K ($z$$<$4.5 kpc) and the PP-Arch ($z$$<$0.9
kpc). All of these show metallicities of 0.5--1 times solar. FUSE data of
extra-galactic targets show \OVI\ absorption at IVC velocities in many cases and
in few of these, the \OVI\ absorption appears to be centered at the same
velocity as the \HI\ of the IVC. However, it is not always clear whether the
\OVI\ is associated with the IVC, or whether the Galactic \OVI\ component is
just broad. For the two IVC sightlines have been studied in detail
(PG\,0804+761\cite{PG0804} and PG\,1259+593\cite{PG1259}) the properties of the
gas (distance, metallicity, ionization structure, the presence of hot gas) are
compatible with the notion that the IVCs are a manifestation of the Galactic
Fountain.

\begin{figure}
\centerline{\vbox{\psfig{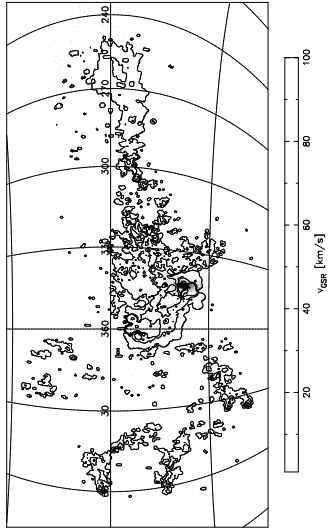}}}
\caption[]{Map of the Magellanic Stream (from \cite{HWS,Morras}). The greyscale
shows velocities, as identified by the wedge. Intensity contours show column
densities of 2, 25, 75 and 350 \tdex{18}\,\cmm2. Dots shows the current particle
positions in the model of Gardiner \& Noguchi\cite{GN}, while solid lines show
the orbits of the LMC and SMC in this model.}
\end{figure}

\section{The Magellanic Stream}
The Magellanic Stream is comparatively well understood. For a while it seemed
possible that the Stream is the result of ram-pressure stripping\cite{MDram},
but since the identification of the leading arm\cite{Lu} the tidal model is
clearly favored. Two measurements of \SII/\HI\ in the Stream exist. Toward
NGC\,3783, ($l$,$b$)=(287\deg,+23\deg), Lu et al.\cite{Lu} found a value of
0.25$\pm$0.07 times solar, while toward Fairall\,9,
($l$,$b$)=(295\deg,$-$58\deg) Gibson et al.\cite{GibMS} find 0.33$\pm$0.05 times
solar. In the most complete published model\cite{GN} the combined tidal force of
the LMC and Galaxy on the SMC peaked 2\,Gyr ago, during their previous
peri-galacticon passage. As a result about 2.5\tdex8\,\Msun\ of material (out of
an original total of 8\tdex8\,\Msun\cite{Stan}) was extracted from the outer
parts of the SMC. One orbit later the MCs are again near perigalacticon, now
followed by a trailing tidal arm (the Magellanic Stream, M=1.5\tdex8\,\Msun) and
a leading arm (M=\dex8\,\Msun). Figure 1 presents a map of the \HI\ in the
Magellanic System, in a coordinate system where galactic longitude $l$=270\deg\
runs along the equator. 

\section{Complexes A and C}
Complex~A is the only HVC with a known distance. A bracket of 4--10\,kpc was
derived by van Woerden et al.\cite{ADUMa}. A FUSE spectrum of the star
PG\,0832+675 improves this to 8--10\,kpc. This implies a gaseous mass of
2\tdex6\,\Msun\ (including a correction for He and assuming a 20\% ionized
fraction). So far, its metallicity is unknown, but FUSE spectra of two
extra-galactic objects suggest a value $<$0.2 times solar, although
contamination by H$_2$ lines is a major problem.
\par The second largest HVC is complex~C. A map of its very complex velocity
field is presented by Wakker\cite{WABS}. Its distance is still fairly uncertain,
but probably between 5 and 20\,kpc, implying a total mass of
3--50\tdex{6}\,\Msun. Its metallicity is comparatively well-known. Initially,
Wakker et al.\cite{MRK290} measured the Sulphur abundance in the direction of
Mrk\,290 (N(\HI) =9\tdex{19}\,\cmm2) as (S/H)=0.09$\pm$0.02$\pm$0.02 times
solar. Here the first error is statistical and the second systematic, and this
value takes into account H$^+$, S$^{+2}$ and \HI\ small-scale structure. Sulphur
is a good element to use in this game, as (a) it is mostly undepleted onto dust
grains, (b) S$^+$ is the dominant ionization stage in the diffuse ISM and (c)
the \SII\,\ll1250, 1253 and 1259 lines are neither too strong nor too weak.
Oxygen is also good, as it's ionization is strongly tied to that of hydrogen,
although some oxygen may be depleted onto dust.
\par Richter et al.\cite{PG1259} measured abundances in the spectrum of
PG\,1259+593 (N(\HI)= 9\tdex{19}\,\cmm2) and found: \NI/\HI=0.012$\pm$0.005,
\OI/\HI=0.09$^{+0.13}_{-0.05}$, \AlII/\HI=0.10$^{+0.11}_{-0.07}$,
\SiII/HI=0.12$^{+0.11}_{-0.05}$, \SII/\HI=0.14$\pm$0.04 and \FeII/\HI=
\par\noindent 0.054$^{+0.032}_{-0.015}$ times solar. On the other hand, Gibson
et al.\cite{GibC} found that \SII/\HI
\par\noindent =0.33$\pm$0.05 toward Mrk\,817 (N(\HI)=3\tdex{19}\,\cmm2) and
\NI/HI=0.039$\pm$0.007 toward Mrk\,876 (N(\HI)= 1--2\tdex{19}\,\cmm2). In the
latter sightline, Murphy et al. \cite{Murphy} also found \FeII/\HI=
0.48$\pm$0.20, although a large ionization correction is expected because N(\HI)
is low. Yet, a factor 10 is difficult to reconcile with the non-detection of
H$\alpha$ emission. Further, toward Mrk\,290 a STIS spectrum shows that in this
sightline \FeII/\HI=0.52$\pm$0.04, while the FUSE spectrum of Mrk\,817 gives
\FeII/\HI=0.19$\pm$0.05 and \OI/\HI=0.19 times solar. These abundances shows
several interesting patterns.
\par (1) They clearly confirm the low metallicity of complex C found by Wakker
et al.\cite{MRK290}. There is some evidence for variations in the metallicity
across the cloud, considering the two different values found for \OI/\HI\ and
\SII/\HI. Measurements in more directions will be necessary to understand this.
\par (2) In both the PG\,1259+593 and the Mrk\,817 sightline the \SII/\OI\ ratio
is $\sim$1.5 times solar, suggesting either that \OI\ is somewhat depleted onto
dust, or that a substantial fraction of the hydrogen is ionized.
\par (3) Two elements that are usually lightly depleted (Al and Si) seem to be
undepleted in complex~C, as the ratios Al/(O,S) and Si/(O,S) are $>$ solar. This
suggests that there is little or no dust present.
\par (4) Iron still appears depleted in three sightlines: \FeII/\SII\ lies in
the range 0.4--0.6 times solar for Mrk\,290, Mrk\,817 and PG\,1259+593. This may
mean that either there is some dust made exclusively of iron particles, or that
the Fe/S ratio in complex~C is intrinsically subsolar.
\par (5) Nitrogen is clearly underabundant in complex C. Toward PG\,1259+593 an
improved FUSE spectrum also allows us to measure the \NII-\l1083 line, which
shows that N(\NI)$\sim$N(\NII). The intrinsic N/O ratio can be approximated by
\NI/\OI$\sim$0.15. This can be compared with value for N/O measured in irregular
galaxies\cite{KS}, in the outer parts of normal spirals\cite{FGW} and in Damped
Ly$\alpha$ Absorbers\cite{LSB}, which shows that only in the latter kind of
object are values found that are as low as in complex~C.
\par Taken together, the abundances and ratios are compatible with the notion
that complex~C consists of gas in which heavy elements were produced only by
Type II SNe, explaining the low Fe and N abundances, the high ratios of
$\alpha$-element abundances (Si, S) over O, as well as the apparent absence of
dust.

\section{\HI\ HVCs in the Local Group?}
Blitz et al.\cite{Blitz} proposed that most HVCs are intra-Local-Group clouds,
assuming (a) that they contain dark matter and (b) have a median distance of
1\,Mpc. Possibly the subset of smaller, compact HVCs forms this population. In
this model each HVC is assumed to be self-gravitating and its distance and mass
can be derived from observables (velocity dispersion, area and flux) using the
virial theorem, assuming a value for $f$, the ratio of the \HI\ mass to the
total mass (\HI+H$^+$+He+dark matter). Blitz et al.\ favor $f$=0.1.
\par Figure 3 shows the resulting \HI\ mass function for the subset of compact
HVCs, for $f$=0.1 (thin dotted line) and $f$=0.01 (thick dotted line). These
distributions are compared to the \HI\ mass function of the Local Group (lower
solid line) and of the general field derived from a deep Arecibo
survey\cite{HIMASS}, scaled to match the Local Group at M(\HI)$>$\dex9\,\Msun\
(upper solid line).
\par Clearly, if the HVCs are virially stable and have $f$=0.1, there would be
10--50 times more massive HVCs in the Local Group than galaxies. There would
also be 50--200 times more than in a comparable volume outside groups. There are
four ways to reconcile this discrepancy. First, the Local Group may be unusual
and indeed have many more massive starless \HI\ clouds than the general field.
Second, the field \HI\ mass function may have been underestimated by a factor
$\sim$10. Third, the value of $f$ may 0.01 or lower (e.g.\ by increasing the
ionization). However, this implies that most distances are in the range
20--200\,kpc, and the HVC ensemble would not fill the Local Group, but rather be
concentrated around the Milky Way. None of these three possibilities seems
likely. Fourth, only a small fraction of the HVC sample may be Local Group
objects. For $f$=0.1 there might be up to 5--10 such objects without causing a
discrepancy. For lower $f$ proportionally more are allowed to exist. 

\begin{figure}
\centerline{\vbox{\psfig{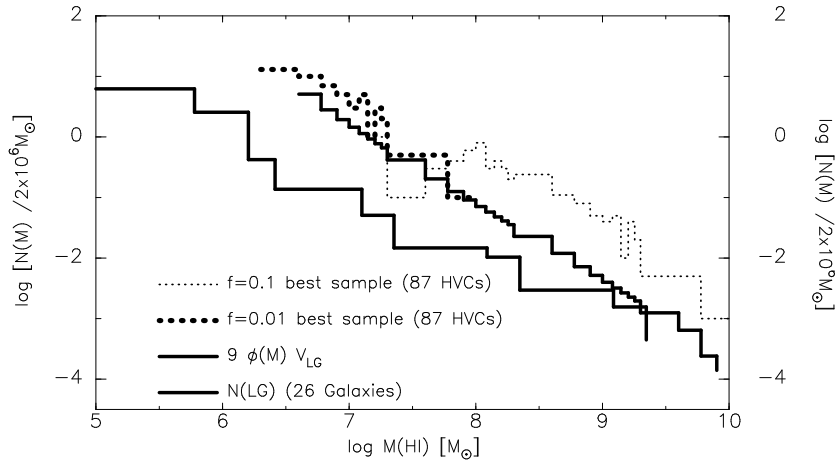}}}
\caption[]{\HI\ mass function of the Local Group (lower solid line), of the
general field (upper solid line, scaled by a factor 9 to match the Local Group
at high masses), and of the HVC sample assuming $f$=0.1 (thin dotted line) and
$f$=0.01 (thick dotted line).}
\end{figure}

\section{High-velocity \OVI}
In its first two years of operation, FUSE has observed 219 extra-galactic
objects. Of the 154 observations that were public as of September 2001, a
reasonable measurement of Galactic \OVI\ absorption can be obtained in 85 cases
(Wakker et al.\ \& Sembach et al., in preparation). In 56 of these sightlines
high-negative or high-positive velocity \OVI\ is found. There is clearly
high-velocity \OVI\ associated with complex~C: in all nine sightlines projected
onto that HVC it is detected, although in three of these the \OVI\ only extends
out to the velocity where the \HI\ peaks. In one sightline through complex~A a
weak \OVI-HVC can be seen. The Magellanic Stream is detected toward Fairall\,9.
In eight other cases a \HI\ HVC lies just a few degrees away and has a velocity
similar to that of the high-velocity \OVI. Twelve sightlines lie in the region
$l$=180\deg--330\deg, $b$$>$35\deg\ and show absorption at velocities
$>$+200\,\kms. Another eleven lie in the region $l$=20\deg--140\deg,
$b$$<$$-$30\deg\ and show absorption at velocities ranging from $-$400 to
$-$150\,\kms\ (six of the sightlines near a \HI\ HVC probably also belong to
this group). Apart from these associations and groupings, there are 14 more
sightlines with high-velocity \OVI. 
\par In the case of complex~C a possible explanation for the associated \OVI\
absorption is that its outer envelope is heated by friction while it moves
closer to the Galaxy. No explicit models of this process exist, however, and it
will require the measurement of other highly-ionized atoms, such as \NV\ and
\CIV\ to discern between the possible explanations.
\par The concentration of twelve sightlines with high-positive \OVI\ lies
diametrically opposed to the concentration of seventeen with high-negative \OVI\
absorption. The latter lies near the direction to M\,31 and other Local Group
galaxies, which may indicate that we are detecting a filament of hot gas in the
Local Group through which the Milky Way is moving. A more detailed analysis of
the velocities and distribution of high-velocity \OVI\ will be needed to confirm
this possibility.

\vfill

\begin{iapbib}{99}{
\bibitem{Arnal}  Arnal E., Bajaja E., Larrarter J., Morras R., P\"oppel
                    W.., 2000, A\&ApS 142, 35
\bibitem{Baj}    Bajaja E., Cappa de Nicolau C.E., Cersosimo J.C., Martin M.C.,
                    Loiseau N., Morras R., Olano C.A., P\"oppel W.G.L., 1985,
                    ApJS 58, 143
\bibitem{Blitz}  Blitz K., Spergel D., Teuben P., Hartmann D., Burton W.,
                  1999, ApJ 514, 818
\bibitem{FGW}    Ferguson A.M.N., Gallager J.S., Wyse R.F., 1998, AJ 116, 673
\bibitem{GN}     Gardiner L.T., Noguchi N., 1996, MNRAS 278, 191
\bibitem{GibMS}  Gibson B.K., Giroux M.L., Penton S.V., Putman M., Stocke J.T.,
                    Shull M.J., 2000, AJ 120, 1830
\bibitem{GibC}   Gibson B.K., Giroux M.L., Penton S.V., Stocke J.T., Shull M.J.,
                    Tumlinson J., 2001, AJ, in press
\bibitem{LDS}    Hartmann D., Burton W.B., 1997, ``Atlas of Galactic Neutral
                    Hydrogen''
\bibitem{HWS}    Hulsbosch A.N.M., Wakker B.P., 1988, A\&ApS 75, 191
\bibitem{KS}     Kobulnicky H.A., Skillman E.D., 1996, ApJ 471, 211
\bibitem{Lu}     Lu L., Sembach K.R., Savage B.D., Wakker B.P., Sargent W.L.W.,
                    Oosterloo T.A., 1998, AJ 115, 162
\bibitem{LSB}    Lu L., Sargent W.L.W., Barlow T.A., 1998, 115, 55
\bibitem{MDram}  Moore B., Davis M., 1994, MNRAS 270, 209
\bibitem{Morras} Morras R., Bajaja E., Arnal E.M., P\"oppel W.G.L., 2000,
                    A\&ApS 142, 25
\bibitem{Murphy} Murphy E.M., Sembach K.R., Gibson B.K., Shull J.M., Savage
                    B.D., Roth K.C., Moos H.W., Green J.C., York D.G., Wakker 
                    B.P., 2000, ApJ 538, L35, 
\bibitem{PG0804} Richter P., Savage B.D., Wakker B.P., Sembach K.R., Kalberla
                    P.M.W., 2001a, ApJ 549, 281
\bibitem{PG1259} Richter P., Sembach K.R., Wakker B.P., Savage B.D., Tripp T.M.,
                    Murphy E.M., Kalberla P.M.W., 2001b, ApJ, Sep 15
\bibitem{Stan}   Stanimirovic S., Staveley-Smith L., Dickey J.M., Sault R.J.,
                    Snowden S.L., 1999, MNRAS 302, 417
\bibitem{ADUMa}  van Woerden H., Schwarz U.J., Peletier R.F., Wakker B.P.,
                    Kalberla P.M.W., 1999, Nature 400, 138
\bibitem{W91}    Wakker B.P., 1991, A\&Ap 250, 499
\bibitem{WABS}   Wakker B.P., 2001, ApJS, Sep 15
\bibitem{MRK290} Wakker B.P., Howk C., Savage B.D., Tufte S.L. Reynolds R.J.,
                    van Woerden H., Schwarz U.J., Peletier R.F., Kalberla
                    P.M.W., 1999, Nature 400, 388
\bibitem{HIMASS} Zwaan M.A., Briggs F.H., Sprayberry D., Sorar E., 1997, ApJ
                    490,173
}
\end{iapbib}
\end{document}